\input harvmac

\def\O{\Omega}
\def\tr{{\rm tr}}
\def\w{\wedge}
\def\s{\sigma}
\def\bom{{\bar \omega}}
\def\hf{{1\over 2}}
\def\diag{{\rm diag}}

\line{\hfill PUPT-1790 }
\line{\hfill {\tt hep-th/yymmddd}}
\vskip 1cm

\Title{}{Zero-Branes on a Compact Orbifold}

\centerline{$\quad$ { Sanjaye Ramgoolam, Daniel Waldram }}
\smallskip
\centerline{{\sl Joseph Henry Laboratories}}
\centerline{{\sl Princeton University}}
\centerline{{\sl Princeton, NJ 08544, U.S.A.}}
\centerline{{\tt ramgoola@puhep1.princeton.edu, 
   waldram@viper.princeton.edu }}

\vskip .3in 

The non-commutative algebra which defines the theory of zero-branes on
$T^4/Z_2$ allows a unified description of moduli spaces associated
with zero-branes, two-branes and four-branes on the orbifold space. Bundles
on a dual space $\hat T^4/Z_2$ play an important role in this
description. We discuss these moduli spaces in the context of
dualities of K3 compactifications, and in terms of properties of
instantons on $T^4$. Zero-branes on the degenerate limits of the
compact orbifold lead to fixed points with six-dimensional scale but not
conformal invariance. We identify some of these in terms of the ADS
dual of the $(0,2)$ theory at large $N$, giving evidence for an
interesting picture of ``where the branes live'' in ADS.

\Date{Revised 8/98: version to appear in {\it JHEP}}

\lref\pmh{ P-M. Ho and Y-S. Wu, 
   ``Noncommutative Gauge Theories in Matrix Theory,'' 
   {\tt hep-th/9801147}.}
\lref\ddg{ D.E. Diaconescu, M. Douglas and J. Gomis, 
   ``Fractional Branes and Wrapped Branes,'' 
   {\it JHEP} {\bf 02} (1998) 013, 
   {\tt hep-th/9712230}.}
\lref\dgm{ M. Douglas and G. Moore, 
   ``D-branes, quivers and ALE instantons,'' 
   {\tt hep-th/9603167}.}
\lref\ptk{ J. Polchinski, 
   ``Tensors from K3 orientifolds,'' 
   {\it Phys. Rev.} {\bf D55} (1997) 6423, 
   {\tt hep-th/9606165}.}
\lref\deg{ M. Douglas, 
   ``Enhanced gauge symmetries in Matrix theory,'' 
   {\it JHEP} {\bf 07} (1997) 004, 
   {\tt hep-th/9612126}.}
\lref\BKV{ M. Bershadsky, Z. Kakushadze and C. Vafa,
   ``String Expansion as Large N Expansion of Gauge Theories,'' 
   {\it Nucl. Phys.} {\bf B523} (1998) 59,
   {\tt hep-th/9803076}.}
\lref\bercor{ D. Berenstein and R. Corrado, 
   ``Matrix Theory on ALE Spaces and Wrapped Membranes,'' 
   {\tt hep-th/9803048}.}
\lref\bfss{ T. Banks, W. Fischler, S. Shenker and L.  Susskind, 
   ``M Theory as a Matrix Model: A conjecture,'' 
   {\it Phys. Rev.} {\bf D55} (1997) 5112, 
   {\tt hep-th/9610043}.}
\lref\wati{ W. Taylor, 
   ``D-brane field theory on compact spaces,'' 
   {\it Phys. Lett.} {\bf B394} (1997) 283, 
   {\tt hep-th/9611202}.}
\lref\berketal{ M. Berkooz, R.Leigh, J. Polchinski, J.Schwarz, 
      N.Seiberg and E.Witten, 
   `` Anomalies, Dualities, and Topology of $D=6$ $N=1$ Superstring Vacua,'' 
   {\it Nucl. Phys.} {\bf B475} (1996) 115, 
   {\tt hep-th/9605184}.}
\lref\aspmor{ P. Aspinwall and D. Morrison,
   ``Point-like instantons on $K3$ orbifolds,'' 
   {\it Nucl. Phys.} {\bf B496} (1997) 149,
   {\tt hep-th/9705104}.}
\lref\mathref{ C. G. Cullen, 
   `` Matrices and Linear transformations, 2nd Ed.''  
   Addison-Wesley, Reading, Mass. 1972, p. 252. }
\lref\gibb{ G. Gibbons, 
   ``Wrapping Branes in Space and Time,'' 
   {\tt hep-th/9803206}.}
\lref\kal{ P. Claus, R. Kallosh and A. Van Proeyen, 
   ``M 5-brane and superconformal $(0,2)$ tensor multiplet in six
       dimensions,'' 
   {\it Nucl. Phys.} {\bf B518} (1998) 117, 
   {\tt hep-th/9711161}.}
\lref\Kt{ P.S. Aspinwall, 
   ``K3 Surfaces and String Duality,'' 
   Lectures given at TASI96, Boulder, Colorado, 
   {\tt hep-th/9611137}.}
\lref\hoz{ K. Hori and Y. Oz, 
   ``F-Theory, T-Duality on K3 Surfaces and $N=2$ Supersymmetric Gauge
     Theories in Four Dimensions,'' 
   {\it Nucl. Phys.} {\bf B501} (1997) 97, 
   {\tt hep-th/9702173}.}
\lref\bsv{ M. Bershadsky, V. Sadov, and C. Vafa,
   ``D-branes and Topological Field Theories,'' 
   {\it Nucl. Phys.} {\bf B463} (1996) 420, 
   {\tt hep-th/9511222}.}
\lref\hull{ C. Hull, 
   ``U-Duality and BPS Spectrum of Super Yang-Mills Theory and
      M-Theory,'' 
   {\tt hep-th/9712075}.}
\lref\witcom{  E. Witten, 
   ``Some comments on string dynamics,'' 
   {\tt hep-th/9507121}.}
\lref\min{ S. Minwalla,
   ``Restrictions Imposed by Superconformal Invariance on Quantum
      Field Theories,'' 
   {\tt hep-th/9712074}.}
\lref\mroz{ M. Rozali, 
   ``Matrix Theory and U-Duality in Seven Dimensions,'' 
   {\it Phys. Lett.} {\bf B400} (1997) 260, 
   {\tt hep-th/9702136}.}
\lref\strom{ A. Strominger, 
   ``Open P-Branes,'' 
   {\it Phys. Lett.} {\bf B383} (1996) 44, 
   {\tt hep-th/9512059}.}
\lref\kac{ V.G. Kac, 
   ``Representations of classical Lie Superalgebras,'' 
   Vol. 676, Springer-Verlag, Berlin, 1978, p.597.}
\lref\aspb{ P.S. Aspinwall, 
   ``Enhanced Gauge Symmetries and K3 Surfaces,'' 
   {\it Phys. Lett.} {\bf 357B} (1995) 141, 
   {\tt hep-th/9507012}.}
\lref\cds{ A. Connes, M. Douglas and A. Schwarz, 
   ``Matrix Theory and Non-commutative geometry: Compactification on tori'' 
   {\it J. High Energy Phys.} {\bf 02} (1998) 003,
   {\tt hep-th/9711162}.}
\lref\am{ P.S. Aspinwall and D.R. Morrison, 
   ``Point-like Instantons and the Spin(32)/Z2 Heterotic String,''
   {\it Nucl. Phys.} {\bf B496} (1997) 149, 
   {\tt hep-th/9705104}.}
\lref\grt{ O. Ganor, S. Ramgoolam and  W. Taylor, 
   ``Branes, Fluxes, and Duality in Matrix Theory,'' 
   {\it Nucl. Phys.} {\bf B492} (1997) 191, 
   {\tt hepth/9611202}.}
\lref\muk{ S. Mukai, 
   ``Symplectic Structure of the Moduli Space of Sheaves on an Abelian
     or K3 surface,'' 
   {\it Invent. Math.} {\bf 77} (1984) 101.}
\lref\malda{ J. Maldacena, 
   ``The Large N limit of superconformal field theories and supergravity,'' 
   {\tt hep-th/9711200}.}
\lref\jbsr{ J. Brodie and S. Ramgoolam,
   ``On Matrix Models of M5-branes,'' 
   {\it Nucl. Phys.} {\bf B521} (1998) 139,
   {\tt hep-th/9711001}.}
\lref\asp{ P. Aspinwall, 
   ``Point-like instantons and the Spin(32)/Z2 Heterotic String,''
   {\it Nucl. Phys.} {\bf B496} (1997) 149, 
   {\tt hep-th/9612108}.}
\lref\gkp{ S. Gubser, I. Klebanov and A. Polyakov,
   ``Gauge theory correlators from noncritical string theory,'' 
   {\it Phys. Lett.} {\bf B428} (1998) 105, 
   {\tt hep-th/9802109}.}
\lref\witads{ E. Witten, 
   ``Anti-de Sitter space and holography,'' 
   {\tt hep-th/9802150}.}
\lref\trev{ A. Giveon, M, Porrati and E. Rabinovici, 
   ``Target space duality in string theory,'' 
   {\it Phys. Rept.} {\bf 244} (1994) 77, 
   {\tt hep-th/9401139}.}
\lref\ztor{ Z. Guralnik and S. Ramgoolam, 
   ``From 0-branes to torons,'' 
   {\it Nucl. Phys.} {\bf B521} (1998) 129,
   {\tt hep-th/9708089}.}
\lref\nahm{ W. Nahm, 
   ``Self-dual monopoles and calorons,'' 
   in Lect. Notes in Physics. 201, eds. G. Denardo {\it et. al.} (1984), 189.}
\lref\hm{ J. Harvey and G. Moore, 
   ``On the Algebras of BPS states,'' 
   {\tt hep-th/9609017}.}
\lref\ghm{ M. Green, J. Harvey, G. Moore, 
   ``I-brane inflow and anomalous couplings on d-branes,'' 
   {\it Class. Quant. Grav.} {\bf 14} (1997) 47, 
   {\tt hep-th/9605033}.}
\lref\witcons{ E. Witten, 
   ``Constraints on Supersymmetry Breaking,'' 
   {\it Nucl. Phys.} {\bf B202} (1982) 253.}
\lref\bss{ T. Banks, N. Seiberg and S. Shenker, 
   { ``Branes from Matrices,''} 
   {\it Nucl. Phys.} {\bf B490} (1997) 91, 
   {\tt hep-th/9612157}.}
\lref\don{ S. Donaldson and P. Kronheimer, 
   ``Geometry of four-manifolds,'' 
   Clarendon Press, Oxford, 1990.}
\lref\witcfth{ E. Witten, 
   ``On the conformal field theory of the Higgs branch,'' 
   {\it J. High Energy Phys.} {\bf 07} (1997) 003,
   {\tt hep-th/9707093}.}
\lref\abkss{ O. Aharony, M. Berkooz, S. Kachru, N.  Seiberg 
   and E. Silverstein, 
   ``Matrix description of interacting theories in 6 dimensions,''
   {\it Adv. Theor. Math. Phys.} {\bf 1} (1998) 148,
   {\tt hep-th/9707079}.}
\lref\jps{ J. Polchinski, 
   ``Scale and conformal invariance in quantum field theory,'' 
   {\it Nucl. Phys.} {\bf B303} (1988) 226.}
\lref\ks{ S. Kachru and E. Silverstein, 
   ``4d conformal field theories and strings on orbifolds,'' 
   {\it Phys. Rev. Lett.} {\bf 80} (1998) 4855,
   {\tt hep-th/9802183}.}
\lref\dgom{ E. Diaconescu and J. Gomis, 
   ``Neveu-Schwarz Five-Branes And Matrix String Theory On K3,''
   {\it Phys. Lett.} {\bf B426} (1998) 287,
   {\tt hep-th/9710124}.}
\lref\sen{ A. Sen, 
   ``Stable Non-BPS Bound states of D-branes,'' 
   {\tt hep-th/9805019}.} 
\lref\grlapi{ B. Greene,  C. Lazaroiu and P. Yi, 
   ``D-particles on $T^4/Z_n$ orbifolds and their resolutions,''
   {\tt hep-th/9807040}.}
\lref\JM{ C. Johnson and R. Myers, 
   ``Aspects of Type IIB Theory on ALE Spaces,''
   {\it Phys. Rev.} {\bf D55} (1997) 6382,
   {\tt hep-th/9610140}.}
\lref\DGM{ M. Douglas, B.  Greene and D. Morrison,  
   ``Orbifold Resolution by D-Branes,'' 
   {\it Nucl. Phys.} {\bf B506} (1997) 84,
   {\tt hep-th/9704151}.}
\lref\G{ B. Greene, 
   ``D-Brane Topology Changing Transitions,''
   {\tt hep-th/9711124.} }
\lref\DG{ M. Douglas and B. Greene, 
   ``Metrics on D-brane Orbifolds,''
   {\it Adv. Theor. Math. Phys.} {\bf 1} (1998) 184,
   {\tt hep-th/9707214}.}
\lref\Muto{ T. Muto, 
   ``D-branes on Orbifolds and Topology Change,''
   {\it Nucl. Phys.} {\bf B521} (1998) 183,
   {\tt hep-th/9711090}.}
\lref\Mohri{ K. Mohri, 
   ``D-Branes and Quotient Singularities of Calabi-Yau Fourfolds,''
   {\tt hep-th/9707012}.}
\lref\dougci{ M. Douglas, 
   ``D-branes and Matrix Theory in Curved Space,''
   {\it Nucl. Phys. Proc. Suppl.} {\bf 68} (1998) 381,
   {\tt hep-th/9707228}.}
\lref\dougcii{ M. Douglas, 
   ``D-Branes in Curved Space,''
   {\it Adv. Theor. Math. Phys.} {\bf 1} (1998) 198,
   {\tt hep-th/9703056}.}

\newsec{Introduction} 

D-branes with worldvolume transverse to ALE orbifolds, $R^4/G$, in
type II string theories have been studied in 
\refs{\dgm,\JM,\DGM,\G,\DG,\Muto,\Mohri,\dougcii,\dougci}, while $0$-branes
on tori $R^p/Z^p$ have been studied in \refs{\bfss,\wati}. The theory is
derived by finding a representation of the discrete group $G$ or the
translation group $Z^p$. The representation space also admits an
action of operators $X$ corresponding to zero brane positions. These
operators satisfy constraints determined by the group $G$ or $Z^p$.
The case of compact orbifolds, $ R^4/{ \Gamma } $, where $\Gamma$ is
the semidirect product of $G $ and $Z^p$, studied in \pmh, has the
novel feature that the translation operators do not commute with the
discrete group. In this paper we study, in detail, the case of
$T^4/Z_2$.

In section 2, we review the theory of zero-branes on a torus $T^p$.
In particular, we recall how to derive a $p+1$ dimensional theory
from a solution of the translation constraints, and how these
theories produce the moduli space $(T^p)^N/S_N$ for $N$ zero-branes.
This field theory can be viewed as a world-volume theory of $Dp$
branes, and the relation between $0$ and $p$ branes is the one
expected from T-duality.
 
In section 3, we review some facts about dualities of $K3$
compactifications of Type IIA string theory, as summarized in \Kt,
focusing on the perturbative orbifold limit.

In section 4, we describe the theory of $N$ pure zero-branes on
$T^4/Z_2$. It is obtained by finding $X$ operators acting in the
regular representation of the constraint algebra, which we call
Rep. I. We check that the moduli space of multi-zero-branes on the
orbifold is correctly reproduced. We also check that the appropriate
enhancement of the Coulomb branch occurs when a zero-brane hits any
of the sixteen fixed points. We give a description of the resulting
theory as a $U(2N)$ gauge theory on $\hat T^4/Z_2$, where $\hat T^4$
is the T-dual of $T^4$ with inverse volume. The spatial $Z_2$
reflection is combined with a non-trivial gauge transformation,
resulting in a non-trivial gauge bundle on the orbifold.
 
In section 5, we study another representation of the algebra of
constraints, which we call Rep. II. We interpret it in terms of $N$
two-branes at a fixed point on $T^4/Z_2$, and check that the moduli
spaces are consistent with this interpretation. We give a description
in terms of $U(N)$ gauge theory on $\hat T^4/Z_2$, where $Z_2$ is
embedded trivially in the gauge group.

In section 6, we study a class of other representations of the
constraint algebra which are characterized algebraically by the
different representatives of the generator of the $Z_2$ group. The
physical interpretation is most simply given in terms of states with 
wrapped membrane and anti-membrane charge at different fixed 
points of $T^4/Z_2$. We check that the Higgs and Coulomb branches 
have the form expected from this physical interpretation.

In section 7, we compare the relation between the theories on
$T^4/Z_2$ and $\hat T^4/Z_2$ to relations we might expect from
dualities of $K3$ compactifications. Some facts about these dualities
are collected in Appendix A.
We write down the 
relevant element of $O(\Gamma_{4,20})$. A lot of its intricate 
structure is deduced from properties 
of the gauge theory description of branes on the two spaces. 
There is a relation to the point-like instantons 
at fixed points of \refs{\berketal,\asp}. 

In section 8, we study relations between the moduli spaces of vacua
of the theories we construct and moduli spaces of instantons on
$T^4$, invariant under certain $Z_2$ symmetry groups.
 
In section 9, we describe the construction of six-dimensional fixed
points by taking certain limits of the parameters of the orbifold
compactification. We give the ADS interpretation of these fixed
points, following the correspondence between $ADS_7$ and the large
$N$ $(0,2)$ theory proposed by Maldacena \malda, and elaborated on by
\gkp\witads.

\newsec{Review of related quotients} 
 
The theory of $N$ zero-branes in flat space is described by the action:
\eqn\bfssact{ L={1\over 2g} \int {dt}
   \,\tr\biggl[ D_t{ X^i}D_t{ X^i} + 2{\theta^T}D_t{\theta} 
      - {1\over 2}[X^i,X^j]^2 
      - 2{\theta^T} \gamma_i [\theta ,X^i] 
   \biggr]}
where $D_t = \partial_t - iA_0$. The general procedure for describing
the zero-branes on a quotient of $R^p$ is to find a representation of
the quotienting group acting on the fields $X$. Let us review this in
two cases.

\subsec{$T^4 = R^4/Z^4 $} 

The theory of $N$ zero-branes on $T^4$ is defined by starting with
the quantum mechanics of zero-branes and imposing constraints.
\eqn\cont{\eqalign{ 
  & U_a X^{b} U_a^{-1} = X^{b} +2 \pi R_a \delta^{ab}, \cr 
  & U_a X^{i} U_a^{-1} = X^{i}.  \cr }} 
Here the $U_a$ represent the four discrete
translations generating the quotienting group $Z^4$, and thus commute
with each other. The $X^a$ are related to positions of the zero-branes
on the torus and $X^i$ are related to transverse positions.
 
The constraints \cont\ can be solved by writing
\eqn\sol{\eqalign{
   & X^a  =  i\partial^a + A^a(x) \cr 
   & X^i = X^i (x) \cr 
   & U_a = e^{ 2\pi i R_a x_a} \cr }} 
where $x^a$ are periodic variables
\eqn\period{
   x^a \sim x^a + { 1 \over R_a}, } 
{\it i.e.} they are coordinates on the dual torus $\hat T^4$. In the
simplest case the gauge fields, and adjoint matter $X^i$ obey periodic
boundary conditions. We have then, a {\it local theory} with {\it
$U(N)$ gauge symmetry } on the T-dual torus $\hat T^4$. This theory
has a moduli space of $(T^4 \times R^5)^N/S_N$ as expected for zero
branes on the space $T^4 \times R^5$.

This $4+1$ dimensional Yang Mills theory can be viewed as the
world-volume theory of $N$ four-branes on $\hat T^4$. The
correspondence between the zero-branes on $T^4$ and the $4$-branes on
$\hat T^4$ is just T-duality on all four circles of the torus, an
element of the T-duality group $O(4,4,Z)$ of the torus, as reviewed
for example in \trev.

More generally one considers non-trivial bundles characterized by
non-zero Chern classes $\tr [ X_{a}, X_{b}]$ and $\tr \epsilon^{abcd}
X_{a} X_{b} X_{c} X_ {d}$, which correspond to systems of zero-branes on
$T^4$, together with extra two- and four-branes respectively \refs{\grt,\bss}.
The appearance of moduli spaces $S^N (T^4) $ in these more general
systems was studied in \ztor. One can also consider a generalization
where the elements corresponding to translations are represented by
operators which only commute up to a phase. These are related to
backgrounds with $B$-fields \cds, and lead to non-commutative gauge
theories.

\subsec{$R^4/Z_2$} 

The quotient by a discrete $Z_2$ is described by:
\eqn\condsq{\eqalign{ 
  & \Omega X^a \Omega = -X^a \cr 
  & \Omega X^i \Omega = X^i, \cr } } 
where $\Omega^2 = 1$. A number of interesting properties of the moduli
spaces corresponding to the coordinates transverse (``Coulomb
branches'') as well as to coordinates parallel (``Higgs branches'') to
the $R^4/Z_2$ ALE space were elucidated in \refs{\dgm,\deg,\ddg,\bercor}.
 
For a single zero-brane, there is a Higgs branch $R^4/Z_2$
corresponding to the zero-brane moving on the ALE. For a generic
point on the Higgs branch, the Coulomb branch is $R^5$. When the zero
brane hits the fixed point, the Coulomb branch is enhanced to
$(R^5)^2$. In order for the ALE space to be a perturbative string
vacuum, there must be $1/2$ a unit of B-field \aspb\ on the collapsed
two-cycle at the singularity. The enhanced Coulomb branch can then
be interpreted in terms of a pair of states which carry opposite membrane 
charge, wrapped on the fixed point, and having a net zero-brane charge, 
because of the coupling
\eqn\cb{  
\int C \wedge (F-B)} 
on the world-volume of the two-brane. We will be more precise 
about the contributions to the zero-brane charge from $F$ and $B$ in 
section four. $C$ is the one-form potential
coupling to a zero-brane.

\newsec{Review of duality on $T^4/Z_2$}

In this paper we will construct the theory of zero-branes on
$T^4/Z_2$. By analogy with the construction on $T^4$ we might expect
the description to be a gauge theory on some dual compact orbifold,
exchanging zero-brane charge for four-brane charge and inverting the
volume of the orbifold. In this section we will briefly review what
kind of duality transformation we might expect.

There are two obvious candidates. A first possibility is to make a
T-duality on the covering torus $T^4$ to a dual torus ${\hat T}^4$ 
(with inverse volume) and then project again to ${\hat
T}^4/Z_2$. Alternatively, one can view $T^4/Z_2$ as an orbifold limit
of a smooth K3 manifold, with sixteen $A_1$ singularities. For $U(N)$
gauge theories with instanton number $k$ on K3, it is known that there
is a generalization of Nahm duality for $T^4$ \nahm, the Fourier-Mukai
(FM) transform~\refs{\muk,\dgm,\hoz,\dgom}, which exchanges $N$ with 
$k-N$.

It is important to note that the two candidate dualities are not the
same. In the case of T-duality inherited from the double cover $T^4$,
for one zero-brane on $T^4/Z_2$ we need two zero-branes on the
$T^4$. Under duality this should turn into two four-branes on ${\hat
T}^4$, described by a $U(2)$ gauge theory. Modding out by $Z_2$ the
theory remains $U(2)$. Thus we see that a single zero-brane on
$T^4/Z_2$ should map to two four-branes on ${\hat T}^4/Z_2$. One
notes further that the dual torus ${\hat T}^4$ should have the inverse
volume of the original $T^4$. Thus if $V$ is the volume of the original
$T^4/Z_2$ then the dual space ${\hat T}^4/Z_2$ must have volume $1/4V$
rather than simply $1/V$.

As for the ALE singularity, the geometrical orbifold is not a
perturbative string background. The perturbative background has
one-half a unit of $B$-field on each collapsed two-cycle~\aspb. Since
we construct the zero-brane theory from perturbative open strings, we
are really concerned with duality of the $T^4/Z_2$ orbifold with
$B$-field on the collapsed cycles.

The Mukai charge $(Q_4,Q_2,Q_0) $ of a system of branes characterizes
$Q_4$ four-branes, described by $U(Q_4)$ gauge theory, with magnetic
flux $Q_2$ giving the number of two-branes, and with zero-brane charge
$Q_0$. $Q_0$ includes a contribution from the instanton number,
measuring the number of physical zero-branes, together with $-Q_4$
induced from the curvature of the $K3$ \refs{\bsv,\hm,\ghm}. Under FM duality,
$Q_4$ and $Q_0$ are exchanged. Thus, in contrast to the T-duality
inherited from $T^4$, a single zero-brane transforms to a system of
{\it one} four-brane with instanton number one

To understand how the volume of the K3 transforms under Fourier-Mukai
duality, one identifies the duality with a particular element of the
duality group of IIA string theory on a K3. The structure of the
string moduli space has been described by Aspinwall~\Kt\ and is
summarized in the Appendix. The relevant point here is that the volume
of the K3 is encoded in a normalized vector $B'$ in the total
cohomology $H^*(K3,R)\cong R^{4,20}$. The integer cohomology forms a
even self-dual lattice $H^*(K3,Z)\cong\Gamma_{4,20}\subset
R^{4,20}$. Points in the lattice represent the Mukai charge of a
system of four-branes, membranes and zero-branes wrapped on the cycles
defined by the point in $\Gamma_{4,20}$. There is a group of discrete
rotation symmetries which leave the lattice invariant and define the
same physical system. These are the dualities of the theory. If we
write $\omega$ as the single element generating $H^0(K3,Z)$ and
$\omega^*$ as the dual element generating $H^4(K3,Z)$, then, if there
is no $B$-field on the K3, the vector defining the volume is given by
\eqn\BprimenoB{
   B' = \omega^* + \alpha\omega } (this vector is always normalized so
that the coefficient of $\omega^*$ is one), and $\alpha$ is the volume
of the K3. Similarly an element of the lattice which describes only a
sum of zero-brane and four-brane Mukai charge is
\eqn\zerofourMukai{
   p = Q_4 \omega^* + Q_0 \omega } One of the dualities which leaves
the $\Gamma_{4,20}$ lattice invariant is the exchange of $\omega$ and
$\omega^*$. Writing $\omega\to\bom=\omega^*$ and
$\omega^*\to\bom^*=\omega$ we find
\eqn\newBpM{\eqalign{
   p &= Q_0 \bom^* + Q_4\bom \cr B' &= \bom^* + \bom/\alpha }} where
we have rescaled $B'$ to get the correct normalization with respect
to $\bom^*$. We see that this duality has realized the
Fourier-Mukai transformation, exchanging the zero- and four-brane
Mukai charges. Since $\alpha\to 1/\alpha$, in contrast to the
T-duality from the torus where $V\to 1/4V$, under Fourier-Mukai
duality we have $V\to 1/V$.

For simplicity we did not include the $B$-field in the above
discussion. This is done in the Appendix. We also note that one
might expect that the duality inherited from the T-duality of $T^4$ is
also realized as a symmetry of the $\Gamma_{4,20}$ lattice. We will
return to these questions later.

\newsec{The Regular Representation (Rep. I)}

\subsec{Algebra, solutions and non-commutativity}

The space $T^4/Z_2$ can be thought of as the plane $R^4$ modded out by
the group $\Gamma$ generated by translations $U_a$ (where $a$ runs
from $1$ to $4$) and an inversion $\O$. These generators satisfy the
conditions
\eqn\algebra{
   \O U_a \O = U_a^{-1}, \qquad \O^2 = 1, } and otherwise commute.

A general element of the quotienting group is given by
\eqn\genstate{
   U_g = \prod_a U_a^{n_a}\O^p, \qquad n_a\in Z, \quad p\in\{0,1\}.  }
This is naturally extended to the group algebra by considering
combinations with complex coefficients.

To write an action for D0-branes on $T^4/Z_2$ we need to find a set of
operators such that
\eqn\ops{\eqalign{
   U_aX^bU_a^{-1} &= X^b + 2\pi R_a {\delta_a}^b \cr \O X^a\O &= - X^a
   \cr U_aX^iU_a^{-1} &= X^i \cr \O X^i\O &= X^i.  }} 
A procedure to
find the solution of these conditions in the regular
representation, for arbitrary constraint algebra constructed from
translations and discrete subgroups of rotations, is given in \pmh.
One introduces a set of dual operators $\tilde{U}_g$ and the
solution for $X^a$ and $X^i$ becomes a general function in the
$\tilde{U}_g$. In our case, one finds that the $\tilde{U}_g$
satisfy the same algebra as the original $U_g$.

\subsec{Explicit representation}

Here we will give an explicit solution of \ops. Let us take
\eqn\explicit{
   \O = \pmatrix{\ast & 0 \cr 0 & -\ast} \qquad 
   U_a = e^{2\pi iR_ax^a}\pmatrix{1 & 0 \cr 0 & 1}, } 
acting on the space of pairs of functions on $\hat T^4$. The operator $*$
acts on functions as $* f(x) = f(-x) *$. We then have
\eqn\expsol{
   X^a = i\partial^a + \pmatrix{A^a & B^a \cr 
     B^{\dag a} & \hat{A}^a} \qquad 
   X^i = \pmatrix{X^i & Y^i \cr Y^{\dag i} & \hat{X}^i} } 
Where $A^a$, $\hat{A}^a$ and
$Y^i$ are odd functions under $x^a\to-x^a$ while $B^a$, $X^i$ and
$\hat{X}^i$ are even. Also $A^a$, $\hat{A}^a$, $X^i$ and
$\hat{X}^i$ are all real. In general, to describe $N$ D0-branes
all these functions are replaced by $N\times N$ matrices.

This corresponds to the regular representation discussed in \pmh.
This representation is a Hilbert space with a distinct basis state for
each element of the group $\Gamma$, i.e.
\eqn\hilbert{
   \big|U_g\big> = U_g\big|0\big> } 
for each $U_g\in\Gamma$.

In the representation above, the vacuum $\big|0\big>$ is identified with the
pair of functions $$\pmatrix {1 \cr 1 \cr }. $$ If we choose the
action
$$\Omega \pmatrix { 1 \cr 1 \cr } = \pmatrix { 1 \cr -1 \cr },$$ 
states in the Hilbert space can be mapped to pairs of functions.

The solution can be expressed in the language of \pmh, 
\eqn\pmsol{\eqalign{ 
   X^a &= A^a (\tilde U^a ,\tilde \Omega) \tilde R + \tilde d^a \cr 
   X^i &= X^i (\tilde U^a ,\tilde \Omega) }} 
where $\tilde \Omega $ and $\tilde U$ generate
the commutant of the algebra of $U$ and $\Omega $ in this Hilbert
space and form an isomorphic algebra. (In the more general
representations to be discussed later, the commutant is not an algebra
generated by such a set of generators.) The operators $\tilde R$ and
$\tilde d$ satisfy $\tilde R f(\tilde U^a , \Omega)= f(\tilde U^a,
-\Omega) \tilde R$, $\tilde R^2=1$ and $\tilde d^a \tilde U^b =
\tilde U^b \tilde d^a + \tilde U^b \tilde R \delta^{ab} R_a$. These
can be expressed as 
\eqn\explicitPM{\eqalign{ 
  & \tilde{\O} = \pmatrix{1 & 0 \cr 0 & -1} \qquad 
  \tilde{U^a} = \pmatrix{\cos(2\pi Rx^a) & i \sin(2\pi Rx^a) \cr 
     i \sin(2\pi Rx^a) & \cos(2\pi Rx^a)} \cr 
  & \tilde{R} = \pmatrix{0 & 1 \cr 1 & 0}. \qquad 
  \tilde d^a = \pmatrix{ i\partial^a & 0 \cr 0 & i \partial^a \cr } }}

Including $A_0$, which has a similar form to $X^i$, we obtain the
action on the dual torus $\hat{T}^4$. We find, in a $U(2)$ invariant
form,
\eqn\taction{
   S = \int_{\hat{T}^4} \left\{
      -{1\over 4g}\tr F_{ab}F^{ab} - {1\over 2g}D_aX^iD^aX_i 
      \right\} } 
where
\eqn\covderF{
   D_aX^i = \partial_a X^i - i[A_a,X^i] } and $F$ is the usual field
strength for $A_a$ considered as a $U(2)$ field.

The action is not in fact invariant under arbitrary $U(2)$ gauge
transformations on $\hat T^4$. Only those gauge transformations which
commute with the constraint ({\it i.e.} with $\O$) are preserved. This means
that a general infinitesimal gauge parameter $\Lambda$ is given by
\eqn\gtransf{
   \Lambda = \pmatrix{\lambda & \mu \cr \mu^{\dag} &
      \hat{\lambda}} } 
where $\lambda$ and $\hat \lambda$ are
even functions while $\mu$ is odd. The condition satisfied by
the gauge transformations is {\it not local } on $ \hat T^4$.

Since all the fields have a definite parity under $x^a\to -x^a$, the
action can be further reduced to one on $\hat{T}^4/Z_2$. We find
\eqn\orbaction{\eqalign{
   S =& \int_{\hat{T}^4/Z_2} \Big\{ - {1\over2g}\tr\left(F - iB\w
B^{\dag}\right)^2 - {1\over2g}\tr\left(\hat{F} - iB^{\dag}\w
B\right)^2 - {1\over g}\tr (DB)(DB^{\dag}) \cr & - {1\over g}\tr
\left(DX-iBY^{\dag}+iYB^{\dag}\right)^2 - {1\over g}\tr
\left(D\hat{X}-iB^{\dag}Y+iY^{\dag}B\right)^2 \cr & - {2\over g}\tr
\left(DY-iB\hat{X}+iXB\right)
\left(DY^{\dag}-iB^{\dag}X+i\hat{X}B^{\dag}\right) \Big\} }} where we
have the covariant derivatives
\eqn\derivs{\eqalign{
   DB &= dB -iA\w B-iB\w \hat{A} \cr DX^i &= dX^i - i[A,X^i] \cr
   D\hat{X}^i &= dX^i - i[\hat{A},\hat{X}^i] \cr DY^i &= dY^i -iAY^i +
   iY^i\hat{A} }} 
We see that $A$ and $\hat{A}$ are gauge fields (in
general $U(N)\times U(N)$). For the scalar matter fields, $X^i$ and
$\hat{X^i}$ are in the adjoint of the left and right $U(N)$
respectively, while $Y^i$ is in the fundamental $({\bf N},\bar{\bf
N})$ of each group. Similarly $B_a$ is a one-form charged in the
fundamental of each group. This structure is similar to the $U(N)
\times U(N)$ gauge symmetry for zero-branes on $R^4/Z_2$.

The theory on $\hat T^4 /Z_2$ is a $U(2N)$ gauge theory, with
boundary conditions which leave a manifest $U(N) \times U(N)$ gauge
symmetry. The massless vector fields are $U(N) \times U(N)$ gauge
fields. The $B^a$ fields appear as massive spin-one matter. That we
can write a local theory on $\hat T^4/Z_2$ is related to the fact
that the fields can be viewed as sections of a non-trivial bundle on
that space. The non-triviality comes from the gauge transformation
by $\sigma_3 \otimes 1$ which accompanies the transition from patches
related by reflection in the fixed points, where $\sigma_i$ are the Pauli
matrices. Strictly speaking the bundle has a singularity at the
fixed points, in the sense that the patching functions fail to obey
the condition $U_{\alpha \beta} U_{\beta \gamma} U_{\gamma \delta }
=1 $ at the fixed points.

The singularity of the bundle structure in the neighbourhood of a
fixed point can be characterized by a Wilson loop on a non-trivial
path in the space with the point removed. This has non-trivial
$\pi_1$ since it is retractable to $S^3/Z_2$ which has $\pi_1
(S^3/Z_2) = Z_2$. Along the lines of \asp\ we expect that this should
be related to the existence of two-brane charge at the collapsing
two-cycle of the blown-up space. In the blown-up space, this would be
realized as $\int \tr F$ on the two-cycle. In this sense the breaking
from $U(2N)$ to $U(N) \times U(N)$ may be understood as a result of
having a non-trivial background at the fixed point. Typically there
would also be a contribution to the instanton number at the fixed
point from this background.

\subsec{Moduli space for one zero-brane }

Since the theory we described above has some rather peculiar
properties, it is useful to demonstrate that it is indeed describing
zero-branes on $T^4/Z_2$. We show here that we do get a Higgs branch
moduli space of $T^4/Z_2$, together with the enhanced Coulomb branches
whenever the zero-brane hits any fixed point.

Since we are looking for zero-energy configurations, it will suffice
to consider configurations of zero field strength $F=0$. Recall how
these give rise to the expected moduli space if we just had the
$U(2)$ theory of two zero-branes on $T^4$. In that case,
the moduli space is described by constant gauge fields $A_a$ and
scalars $X^i$. We can simultaneously diagonalize the $U(2)$ matrices, and
allowing for the residual $Z_2$ symmetry, we have a moduli space of
$(T^4\times R^5)^2/Z_2$. This is the configuration space of two
identical zero-branes on the space $T^4 \times R^5$.

In this case the reduced gauge symmetry means that we cannot
diagonalize the $U(2)$ matrices. Consider constant gauge connection.
Since $A$ and $\hat{A}$ are odd functions this means
\eqn\zmA{
   A^a = \pmatrix{0&B^a\cr\bar{B}^a&0} = B^a_1\s_1 +
       B^a_2\s_2 } 
Similarly the allowed constant gauge transformations are
\eqn\zmgauge{
   \Lambda = \pmatrix{\lambda&0\cr0&\hat{\lambda}} =
       \lambda_0 1 + \lambda_3\s_3 } 
These are not enough to
diagonalize $A^a$, at best we can rotate all the field into,
say $\s_1$ so that
\eqn\redzmA{
   A^a = B^a_1\s_1 } 
and there is still a residual gauge symmetry
$\sigma_3$ which identifies $B^a_1$ with $-B^a_1$.

Now we can also identify $B^a_1$ with $B^a_1+2\pi R^a$ (a translation
on the original torus) since these configurations are related by gauge
transformations with
\eqn\largegauge{
   g = \exp(2\pi iR_ax^a\s_1) } (Note that the exponent is an odd
function of $x^a$ and so this is an allowed gauge transformation.)
This, together with the residual gauge symmetry $\sigma_3$, leads to a
Higgs branch moduli of $T^4/Z_2$.

Now we turn to the $X^i$ sector. We need a solution where
\eqn\xcond{
   \partial_a X^i - i[A_a,X^i] = 0, \qquad [X^i,X^j] = 0 } 
where $A^a=B^a_1\s_1$. We can find solutions by the following 
procedure. For $A_a=0$ we clearly have the solution
\eqn\basicsol{
   X^i = X^i_0 1 + X^i_3 \s_3 } 
where $X^i_0$ and $X^i_3$ are constant
(which is allowed given the odd/even properties $X^i$). In such a case
the transverse moduli space is $R^5\times R^5$. We can grow a
solution with $A_a\neq 0$ by making a gauge transformation on the
covering $R^4$ of the form (which has the correct odd/even properties)
\eqn\largegauge{
   g = \exp(iB^a_1x_a\s_1) } 
giving $A^a=B^a_1\s_1$. In general such a
gauge transformation on $R^4$ is not a valid gauge transformation on
$\hat T^4$ since it fails to be single valued on $\hat{T}^4$. However,
we can use it to generate solutions provided the resulting fields are
properly single valued.

To check single-valuedness, we write the transformed $X^i$ :
\eqn\newX{
   X^i = X^i_0 1 + X^i_3\exp(2iB^a_1x_a\s_1)\s_3 } For general $B^a_1$
this is only single valued if $X^i_3=0$. Thus for generic points in
$T^4/Z_2$ the transverse moduli space is $R^5$. The fixed points of
the orbifold correspond to $B^a_1= \pi k^a$, where $k^a$ is a vector
with components $0$ or $R_a$. We then see that at the fixed points,
there is no constraint on $X^i_3=0$ and the moduli space becomes
$R^5\times R^5$.

The moduli space may be viewed as a fibration over the Higgs branch of
$T^4/Z_2$. The fibre at a generic point is $R^5$. However at the
orbifold points, the fibre becomes $R^5\times R^5$. Note that this
is $R^5 \times R^5$ rather than $(R^5 \times R^5) /Z_2$, which is
consistent with the interpretation that the zero-brane decays into 
two distinct objects, free to move only transverse to the fixed point. 
Inspection of the mass formula in the appendix, shows that 
the four states with Mukai vectors $\gamma_{16}$, $-\gamma_{16}$, 
$\gamma_{16}-\omega$ and $-\gamma_{16}+\omega$, are degenerate and are 
the lightest states carrying membrane charge on the collapsed 
cycle $\gamma_{16}$. Each state has half the mass of a single zero-brane.
Taking into account the $B$ field, these states 
carry total charge $(q_2^{(16)},q_0)$ of $(1,\hf)$, $(-1,-\hf)$, 
$(1,-\hf)$ and $(-1,\hf)$ respectively. Thus a zero-brane can decay into 
the pair $(1,\hf)$ and $(-1,\hf)$ with the same charge and mass. In the 
rest of the paper, we will refer to the 
state $(1,\hf)$ as ``membrane'' and the state $(-1,\hf)$ 
as ``antimembrane''\foot{In terms of the Mukai vectors, we note 
that the $(-1,-\hf)$ state is more strictly the pure anti-membrane 
state. However since in what follows we will only be interested in 
the state $(-1,\hf)$, which actually has Mukai vector 
$-\gamma_{16}+\omega$, for simplicity, we will refer to this latter 
state as the ``antimembrane''.}.

\subsec{$S^N(T^4/Z_2)$ for $N$ zero-branes}
 
The discussion of the Higgs branch where $X^a \ne 0$ and $X^{i} = 0 $
generalizes in a simple way. As before we can get the expected moduli
space by restricting attention to zero-field-strength configurations.
(We expect that the condition of zero energy will force the $X^a$ to be
space-time independent up to gauge transformations, along the lines of
\witcons.)

Any constant matrix $X^a$ of the form
$$\pmatrix { 0 & A \cr A^{ \dagger } & 0 \cr }$$ can be put in a
standard form
\eqn\stdfrm{\eqalign{ \pmatrix { 0 & D \cr D & 0 \cr } }} 
where $D$ is diagonal and real, using
gauge transformations of the form $$ \pmatrix { U & 0 \cr 0 &
V. \cr } $$ This follows from using the Polar Decomposition theorem
\mathref. Then any other matrix of the form 
$$ \pmatrix { 0 & B \cr B^{ \dagger} & 0 \cr } $$ 
which commutes
with $X^a$ can simultaneously be put in the form \stdfrm. Permutation
matrices of the form
$$\pmatrix { \sigma &0 \cr 0 & \sigma \cr }$$ and the matrix
$\diag (1,-1)$ lead to identifications which leave a moduli space with
a Higgs branch $(T^4/Z_2)^N/S_N$. This is as expected for a system of
$N$ zero-branes on $T^4/Z_2$.

\newsec{Rep. II --- Interpretations on $T^4/Z_2$ and $\hat T^4/Z_2$ } 

In this section we study Rep. II given by the following solution of
\ops\
\eqn\algsolv{\eqalign{
  & U_a = e^{ 2 \pi i R_a x_a} \cr 
  & \Omega = * \cr 
  & X_{a} = i D_a(x) \cr 
  & X_{i } = X_{i} (x), \cr }} 
which is even simpler than Rep. I.
The $\Omega $ constraint requires that the gauge field $A_a(x)$ is
odd, the scalar $X_{i} (x)$ is even, and the gauge parameter is even.
The gauge symmetry is $U(1)$ in the simplest case. One generalizes to
$U(N)$ by replacing $\Omega = * $ by $\Omega = * 1$, where $1$ is the
identity matrix in $U(N)$, and $X_m$ by covariant derivatives of
$U(N)$ etc. The theory can be written as $U(N)$ gauge theory on $\hat
T^4/Z_2$.
\eqn\theot{
   S = \int_{\hat{T}^4/Z_2} \left\{-{1\over 4g}\tr F_{ab}F^{ab} -
      {1\over 2g}D_aX^iD^aX_i \right\} } 
There is no non-trivial gauge
transformation accompanying reflection in the fixed points, so the
fields are sections of a trivial bundle.

\subsec{The interpretation on $T^4/Z_2$} 

This representation can be interpreted in terms of a two-brane wrapped
at the origin in $T^4/Z_2$. This is inspired by the interpretation of
a similar representation in the case of $R^4/Z_2$ in terms of a
two-brane at the fixed point \refs{\deg,\ptk}. More generally we have $N$
two-branes at the origin. This is not the only possible interpretation
but appears to be the simplest one consistent with the moduli spaces
of vacua. Because of the B-field at the fixed point, the two-brane
induces a zero-brane charge of $1/2$ via the coupling \cb.

An obvious check of this interpretation is the appearance of the
expected $U(N)$ gauge symmetry of $N$ identical objects. Further
evidence for this picture will be given in section six, where we
describe representations of the algebra \ops\ which correspond to
two-branes at other fixed points than the origin.

\subsec{The moduli space} 
 
The moduli space has no Higgs branch. The constraint that $A$ is odd
rules out the constant connections which, in Rep. I, describe a
$T^4/Z_2$. This is consistent with the picture of a two-brane stuck at
the fixed point.

There is a Coulomb branch where $X_i$ acquire constant expectation
values, and we have a space $R^5$, or more generally $(R^5)^N/S_N$.
This corresponds to the $N$ two-branes moving in the transverse space.

\subsec{Interpretation on $\hat T^4/Z_2$}

We have described above the interpretation of the theory in terms of
branes on $T^4/Z_2$. The form of the theory is also consistent with a
description as a theory of $N$ four-branes wrapped on $\hat T^4 /Z_2$.
Since the bundle is trivial, this is a quotient of $N$ four-branes
wrapped on $\hat T^4 $ with the $Z_2$ action on space only, unlike
Rep. I where the $Z_2$ is embedded non-trivially in the gauge group.

The moduli space is consistent with the picture that this
representation describes a physical four-brane living on $\hat T^4/Z_2$,
and Mukai charge $(N,0,-N)$, with $B= -{1\over 4} \sum\gamma_i$
where $\gamma_i $ are elements of $H^2 (K3)$ corresponding to the
two-cycles collapsed at the fixed points. The Coulomb branch of
$(R^5)^N/S_N$ is interpreted as the configuration space of positions
of the four-branes in the transverse space.

\newsec{General Reps --- Branes at collapsed and smooth cycles} 

In general, we expect that there are representaions of the constraint
algebra corresponding to different combinations of zero-branes, two-branes
and four-branes on $T^4/Z_2$. In this section we will discuss the form of
these generalized reps.

\subsec{Branes at collapsed cycles --- the interpretation}
 
Since $ \Omega^2 = 1$, a simple class of generalizations of Rep. I and
Rep. II is given by choosing $\Omega = * \diag (1 \cdots 1, -1 \cdots
-1)$, with $k$ $(+1)$ eigenvalues, and $l$ $(-1)$ eigenvalues. The
translations are still given by $U_a = e^{ 2 \pi i R_a x^a } $. This
is interpreted in terms of $k$ states with charge $(1,-\hf)$ 
and $l$ states with charge $(-1,\hf)$ at the
fixed point at $0$. Representations which correspond to two-branes
wrapped at different fixed points are obtained by translating by half
a lattice spacing in any direction. We note that $e^{-i \pi k_a x^a }
* e^{ i \pi k_a x^a }= * e^{ 2i \pi k_a x^a }$, so that reps
corresponding to branes at different fixed points are simply obtained
by replacing one or more $1$'s along the diagonal by $e^{ 2 i \pi k_a
x^a }$.

\subsec{Branes at collapsed cycles --- tests of interpretation} 

We will describe tests of the above interpretation using a few
examples.
 
1) Consider first the theory of two membranes at a fixed point of
   $T^4/Z_2$. We expect $U(2)$ gauge symmetry for two identical
   branes, We expect no Higgs branch since the membranes are not free
   to move off the fixed point, and a Coulomb branch of $
   (R^5)^2/Z_2$. All these conditions are satisfied by the proposed
   theory which is just Rep. II with $N=2$. This is a special case
   of the discussion in section 5.
 
2) A system of membrane and antimembrane at the origin has the same
   charges as a zero-brane, and was discussed as the $N=1$ case in
   Rep. I. The pair can split off as a zero-brane which can wander on
   the $T^4/Z_2$, so there is a Higgs branch. When the zero-brane is
   at a generic point on the orbifold, we have a Coulomb branch of
   $R^5$. When it hits any of the fixed points, we get an enhanced
   Coulomb branch $(R^5)^2$.

3) Now consider two membranes and an antimembrane at the origin. This
   is described by choosing $\Omega = * \diag (1,1,-1) $,
   which we will call Rep. III. Analyzing the moduli space, one finds
   that there is a Higgs branch of $T^4/Z_2$, corresponding to a
   membrane-antimembrane pair forming a zero-brane and moving off the
   fixed point. At the origin of the Higgs branch, there is a Coulomb
   branch of $ (R^5)^2/Z_2 \times R^5$. At a generic point of the
   Higgs branch there is $ (R^5)^2$. At a fixed point other than the
   origin, we expect $ (R^5)^3$ (without the $Z_2$ quotient).

   The origin of $T^4/Z_2$ corresponds to the trivial flat
   connection. We can pick a gauge where the points away from the
   origin correspond to connections of the form $$ \pmatrix{ 0& 0
   &B^a \cr 0& 0 &0 \cr B^a & 0 &0} $$ This connection
   corresponding to the point at $B^a$ on $T^4/Z_2$ is obtained by a
   ``gauge-like'' transformation from the trivial connection with gauge
   parameter, $U = e^{ix_a B^a (E_{13} + E_{31}) } $, where $E_{13}
   + E_{31} $ is the matrix with $1$ in the $(13)$ slot and in the
   $(31)$ slot. It is not a real gauge transformation on $\hat T^4$
   because it is not periodic. If we act with this gauge
   transformation on a configuration of the transverse coordinates
   $X^i$, with non-zero components along the diagonal, the resulting
   configuration does not satisfy periodic boundary conditions for
   generic $B^a$. Only if $X^{i}_{33} = - X^{i}_{11}$, are the
   boundary conditions are satisfied. For $B^a = \pi k^a $, we do not
   need this condition for periodicity of the $X^i$, so the moduli
   space is enhanced to $(R^5)^3 $. There is no $Z_2$ modding because
   conjugating the permutation which exchanges the first and second
   entry by $ e^{ix_aB^a (E_{13} + E_{31}) }$ does not give a valid
   gauge transformation for $B^a = \pi k^a $. For multiples of $2 \pi
   R_a$ of course we do get valid gauge transformations, and we
   recover the same Coulomb branch as at the origin of the $T^4/Z_2$.

4) A final example is one where we have $ \Omega = * \diag (1, e^{2
   \pi i R_1 x^1 }, -e^{ 2 \pi i R_1 x^1}) $. We will call this
   Rep. IV. This describes a membrane at the origin and a
   membrane-antimembrane pair at the fixed point $(\pi R_1, 0,0,0)$.
   The membrane-antimembrane pair should be able to split off as a
   zero-brane so we expect a Higgs branch of $T^4/Z_2$. When the zero
   brane is at $(\pi R_1,0,0,0)$ (corresponding to the trivial
   connection) we have an enhanced Coulomb branch of $(R^5)^3$. At
   the non-zero connection corresponding to the zero-brane hitting the
   origin, we have an enhanced Coulomb branch of $ (R^5)^2/Z_2 \times
   R^5$.

These features can be obtained essentially by conjugating the
solutions from before. We can relate solutions of Rep. III to those
of Rep. IV by observing that the matrix 
\eqn\matrns{ 
   M = \pmatrix{1 & 0 & 0 \cr 
        0 & {1\over 2}(1 + e^{2\pi i R_1 x_1}) & 
            {1\over 2}(1 - e^{ 2 \pi i R_1 x_1}) \cr 
        0 & {1\over 2}(1 - e^{2 \pi i R_1 x_1}) & 
            {1\over 2}(1 + e^{2 \pi i R_1 x_1}) } }
conjugates Rep. III into Rep. IV. So the trivial connection in
Rep. III, with its associated Coulomb branch of $(R^5)^2/Z_2 \times
R^5 $, maps to a non-zero connection in Rep. IV. The connection
corresponding to the fixed point $ (\pi R_1,0,0,0)$ maps to the
trivial connection in Rep. IV, again giving the desired Coulomb
branch.

\subsec{Branes at uncollapsed cycles} 

There is a natural generalization of the representations considered
above. We considered above a class of representations where $\Omega =
* T$, with $T^2 =1$, and the gauge fields were restricted to be
periodic or anti-periodic. We now show that each such choice of
$\Omega$ admits generalizations which are characterized by quantities
$\epsilon^{abcd} \tr (X_a X_b X_c X_d)$ and $\tr [X_{a }, X_{b} ]$.
In the absence of $B$ field on the cycles of the torus, these are the
Chern characters which are integral.

We will illustrate by considering the case $\Omega = * T $, where $T$
is a constant matrix squaring to $1$. $X_a$ are still covariant
derivatives, now acting on the space of sections of a non-trivial
$U(2N)$ bundle on $ \hat T^4$ with transition functions constrained
by the form of $\Omega$.

To be more concrete we specify some of the constraints that the
patching functions must satisfy. The $\hat T^4$ is being realized as
a quotient of $R^4$ by 4 translations. There is also a $Z_2$ quotient
to give $\hat T^4/Z_2$. We can give the bundle on the orbifold as a
quotient of a bundle on $R^4$. The $4$ translations are accompanied
by gauge transformations $\Omega_a (x) $. The reflection in the
origin is accompanied by a gauge transformation $T$. Reflection in
other fixed points of the $\hat T^4/Z_2$ are accompanied by gauge
transformations constructed from $\Omega_a$ and $T$. For example a
reflection around the fixed point at $(\pi R_1,0,0,0)$ is accompanied
by $\Omega_1 T$. Patching functions and gauge transformations, viewed
as sections on $R^4$ obey the conditions:
\eqn\condrf{\eqalign{ 
   & g (x + 2 \pi R_a) = \Omega_a g(x) \Omega_a^{-1} \cr 
   & g (-x) = T g(x) T^{-1} \cr }} 
The operator $\Omega $ is a map between the fibre at $x$ and the fibre at
$-x$ satisfying $\Omega \Omega_a \Omega = \Omega_a^{-1}$. The $\Omega_a$ 
allow non-trivial Chern Classes of the $\hat T^4$ bundle. $T$ encodes 
non-trivial bundle structure associated with the fixed points 
of $\hat T^4/Z_2$.

\subsec{Traces and branes} 

We discuss in this section the relation between traces of operators
appearing in \ops\ in these representations of the algebra \algebra\
and brane charges on $T^4/Z_2$. For simplicity we restrict to the
case where the $B$-field is zero on the cycles of the torus.

By analogy with the torus case, the operators $\epsilon^{abcd} \tr
(X_a X_b X_c X_d)$, $\tr [X_a,X_b]$ and $\tr 1 $ measure four-brane
charge, two-brane charge on the uncollapsed cycles of $T^4/Z_2$, and
zero-brane charge. As the above examples suggest, the two-brane charges
at the fixed points $\pi \epsilon_a R_a$ are measured by $ \tr(T
\prod_a U_a^{\epsilon_a})$ where $\epsilon_a$ is $0$ or $1$. There
are some subtleties in the exact map between the traces and brane
charges that remain to be worked out. We expect from the example of
Rep. II (but have not given a general argument) that the zero-brane
charge actually includes, in the presence of two-branes on the
collapsed cycles, the induced charge due to the B-field according to
\cb. It remains to be determined whether, in the presence of
four-branes, it includes the induced charge due to the curvature of the
orbifold.

It is also natural to ask for the interpretation of the traces in
terms of the brane charges on $\hat T^4/Z_2$. This would require a
more systematic understanding of the relation between the theory on
$T^4/Z_2$ and the one on $\hat T^4/Z_2$. We will turn to this in the
next section.

\newsec{Identifying the duality} 

We have seen in section 3 that T-duality of $T^4$ should lead to a map
between the theory on an orbifold $T^4/Z_2$ of volume $V$ and a theory
on $ { \hat T^4}/Z_2 $ of volume $1/{ 4V}$. There is also the
Fourier-Mukai (FM) transformation which exchanges four-brane and
zero-brane Mukai charges and, at least without the presence of
$B$-field, inverts the volume of the orbifold.

The above discussion of various representations of the algebra
\ops\ always culminates in a theory of four-branes on $\hat
T^4/Z_2$. It may also contain other brane charges. In particular, in
Rep. I the appearance of a non-trivial gauge bundle due to the
embedding of the $Z_2$ symmetry in the gauge group suggests there are
membrane and possibly zero-brane charges at the fixed points. 
From the moduli space, we have argued that the different
representations correspond to different collections of zero-branes on
the original $T^4/Z_2$, together with membranes on the collapsed
cycles at the fixed points. It is natural to ask if there is a K3
duality which realizes the transformation between a theory of
zero-branes and membranes on $T^4/Z_2$ to one of four-branes and
perhaps other brane charges on the dual $\hat T^4/Z_2$. 

It is clear that the duality is not Fourier-Mukai. This is apparent
from the fact that Rep. I which describes a single zero-brane on
$T^4/Z_2$ is related to a $U(2) $ gauge theory on the dual space. This
means that a zero-brane is dual to {\it two} four-branes, with possibly
extra two-brane and zero-brane charges. Furthermore, we find that the
volume of the dual orbifold is ${\bar V}=1/{4V}$. Without $B$-field
Fourier-Mukai duality sets ${\bar V}=1/V$. With $B$-field the
situation is worse. As discussed in Appendix A, the volume is not in
general inverted. Furthermore the $B$-field at the fixed point is
rescaled, so the dual orbifold is not a perturbative string
background. We will, however return to this duality in the next
section, combining it with some extra assumptions to learn something
about instantons on $T^4$.

The fact that the zero-brane maps to two four-branes strongly suggests
that the duality in question is that inherited from the action of
T-duality of $T^4$. As discussed in section 3 and Appendix A, for a
system with Mukai charge $(Q_4, Q_2, Q_0)$ we can associate a vector in
the lattice $\Gamma^{4,20}$. Dualities are discrete rotations $O$
which preserve the lattice. From Rep. I we have that one zero-brane
maps to two four-branes and perhaps additional membrane and zero-brane
charges at the fixed points. As discussed in Appendix A, requiring
that the transformation inverts the volume and preserves the $B$-field
flux at each fixed point implies that 
\eqn\constbom{
   \omega = C\left\{2 \bom^* + 2 \bom 
       - \hf \left({\bar\gamma}_1+\cdots+{\bar\gamma}_{16}\right)
       \right\}
}
for some integer $C$, where $\bar\gamma_i$ are the sixteen collapsed
two-cycles. Fractional amounts of $\bar\gamma_i$ are allowed
because they span the Kummer lattice rather than $\Gamma^{4,20}$. From
Rep. I we expect that $C=1$. One can then check that this
transformation preserves the mass of the zero-brane according to the
mass formula in Appendix A, with $C=1$, and $V \rightarrow 1/4V$. This
expression satisfies the desired condition
$\omega\cdot\omega=0$. Another promising aspect of this
transformation is that it mixes zero-brane and collapsed
membrane charge, which is suggested by the non-trivial bundle
structure of Rep. I. 

The amount of zero-brane charge appearing on the dual orbifold is
interesting. With $C=1$, a single zero-brane transforms into two
four-branes and 2 units of zero-brane charge. However, this is Mukai
charge (see section 3 and Appendix). Since the curvature of the
orbifold induces $-2$ units, the actual number of zero-branes is
$4$. This corresponds to $1/4$ of a zero-brane charge at each fixed
point. Fractionally charged instantons at ADE singularities have been
discussed in \refs{\asp,\berketal}. The $U(1)$ instanton constructed in
\berketal\ has membrane charge specified by $\int_{C_i}F/2\pi=1$, or
equivalently $Q_2=-\hf\gamma_i$, and instanton number $-1/4$.
 Following the correspondence discussed in the Appendix,
this is physical zero-brane charge $1/4$. This means that the dual of
the zero-brane following from \constbom\ is consistent with explicit
constructions of instantons at the fixed points. 

Let us now turn to Rep. II. We have argued that this corresponds to a
collapsed membrane at the origin of $T^4/Z_2$. It gives a trivial
bundle on ${\hat T}^4/Z_2$, which is naturally interpreted as single
physical four-brane, which, because of the curvature of the orbifold,
has Mukai vector $(1,0,-1)$. Thus, if $\gamma_{16}$ is the collapsed
two-cycle at the origin, we have
\eqn\gammsix{
   \gamma_{16} = \pm \left(\bom^* - \bom \right),  }
since these arguments cannot fix the overall sign. What about
membranes at other collapsed cycles? We argued in section six, that a
membrane at the fixed point $\pi k^a$ where the components of $k^a$
are $0$ or $R_a$, is described by choosing
$\Omega=*e^{2i\pi k_ax^a}$. If a vector with components 
$q^a$ taking values $ {1/2 R_a }$ or zero, 
gives the coordinates of
 the fixed points in ${\hat T}^4/Z_2$, this means there is a
non-trivial bundle structure around $q^a$ if $2k_aq^a$ is
odd. One finds that this implies that half the fixed points of ${\hat
T}^4/Z_2$ have trivial bundle structure and half have non-trivial
structure. Thus we expect that a membrane at a collapsed cycle other
than the origin is mapped to a single four-brane together with
non-trivial membrane charge at eight of the collapsed cycles of ${\hat
T}^4/Z_2$, and, perhaps, some zero-brane charge. Consider a
membrane at a fixed point (labelled by an index $i$ running 
from $1$ to $15$) with coordinates 
 $\pi k^a$ in $T^4/Z_2$, with $k^a$ not all zero. 
The set of 
fixed points of ${\hat T}^4/Z_2$ with 
non-trivial bundle structure are
characterized by 
\eqn\Ddef{
   \bar D_i = \sum_j {\xi_i}^j \bar\gamma_j }
Here, if the $j$th fixed point in ${\hat T}^4/Z_2$ has coordinates
$q^a$, then ${\xi_i}^j$ is zero or one depending on whether $ 2k_aq^a$ is
even or odd. One then notes that in fact $\hf\bar D_i$ is an element of
the $\Gamma^{4,20}$ 
lattice, since the $\bar\gamma_i$ span the Kummer lattice rather
than $\Gamma^{4,20}$. This suggests that membranes at fixed points
other than the origin transform as
\eqn\othergamm{
   \gamma_i = \pm \left(\bom^* + \bom - \hf \bar D_i\right) }
which, as required is orthogonal to $\gamma_{16}$ and $\omega$, and
squares to $-2$. As with the expression for $\omega$, we see that each
membrane $\hf\bar\gamma_i$ in ${\hat T}^4/Z_2$ leads to $1/4$ unit of
physical zero-brane charge, so that the expression is consistent with 
the explicit constructions of instantons at the fixed
points~\refs{\asp,\berketal}. 

Thus the structure of the gauge theory for zero-branes
and membranes at different collapsed cycles provides much of the
information necessary to find the relevant duality. One finds the full
transformation is given by
\eqn\duality{\eqalign{
   \omega &= 2 \bom^* + 2 \bom 
       - \hf \left({\bar\gamma}_1 + \cdots +
            {\bar\gamma}_{15} + {\bar\gamma}_{16}\right) \cr
   \omega^* &= 2 \bom^* + 2 \bom 
       - \hf \left({\bar\gamma}_1 + \cdots + 
            {\bar\gamma}_{15} - {\bar\gamma}_{16}\right) \cr
   \gamma_i &= \bom^* + \bom - \hf \bar D_i 
       \qquad \qquad i=1,\ldots,15 \cr
   \gamma_{16} &= \bom^* - \bom
}}
This indeed preserves the lattice and maps $T^4/Z_2$ to ${\hat
T^4}/Z_2$ with volume $1/4V$ and preserves the $B$-field flux on
the collapsed cycles. To show this it is useful to note that
\eqn\Drel{
   \bar D_i\cdot \bar D_j = - 8 - 8 \delta_{ij} \qquad
   \bar\gamma_{16}\cdot \bar D_i = 0 }
One notes that the transformation \duality\ singles out one of the
collapsed cycles $\gamma_{16}$, corresponding to the origin. Clearly
there are really sixteen such dualities, each singling out a different
fixed point. Here one particular duality was relevant because
we chose an origin in solving the constraints.

\newsec{Fourier-Mukai duality on $K3$ and point-like instantons on $T^4$} 

While the FM duality does not seem 
to be the right duality for relating the 
theory on $T^4/Z_2$ and $\hat T^4/Z_2$, 
it does allow us, if we make some further 
assumptions, to learn something 
about the moduli space of $Z_2$ invariant 
instantons on $T^4$.  
 
For $k$ zero-branes we have 
a moduli space $(T^4/Z_2)^k/S_k$ which appears 
as a space of vacua in the theory constructed using Rep. I. 
This is dual, by FM duality, to 
$k$ four-branes with $k$ instantons which is 
the theory constructed by representing 
the $Z_2$ by $1$ in the gauge group as in Rep. II.
This duality will actually change the $B$-field at 
the fixed point. If we assume, however, that 
the topology of the  
moduli space is not changed by shifting the $B$-field
or the volume from one generic non-zero value to another, we can 
predict some properties of moduli spaces
of instantons in theories constructed along the lines of Rep. II.  
We are led to look for configurations of instanton
number $2k$ on $T^4$,  
which are invariant under the $Z_2$  
projection. A subset of such instantons 
will be point-like. Their moduli space
should not be larger than $(T^4/Z_2)^k/S_k$. 
We can, in fact, construct the symmetric product space 
by looking at $k$ pairs of point-like instantons 
placed in a $Z_2$ invariant configuration on $T^4$. 
It follows then, that for this choice of instanton number, 
there are no $Z_2$ symmetric fat instantons. 
It might appear that 
the moduli space is actually larger since 
we could choose different 
embeddings of the point-like instantons 
in the $U(k)$. However it is known that 
on $R^4$, the Donaldson-Uhlenbeck compactification \don\ 
does not keep track of the group orientations of the 
point-like instantons. (This 
has an interpretation in the context of 
the $(0,2)$ theory \jbsr.)
It is reasonable to conjecture then 
that this is also true for $T^4$. If we did 
keep track of the embedding of the instanton, 
there would be a subgroup of the gauge group 
which commutes with it, which could be measured
by a non-trivial Wilson loop in the space 
with the point removed. Since $\pi_1 (S^3)$ 
is trivial, there is no possibility of such non-trivial 
Wilson loop \asp. 

To get precise agreement with the symmetric product spaces, 
we would need to prescribe more carefully what kind of point-like 
instantons are allowed at the fixed points. 
The prescription which would work is to keep 
only those which can be obtained
by taking a limit where point-like instantons on $T^4/Z_2$ 
approach from a smooth point to the 
fixed point. However there may be more general 
point-like instantons at the fixed points, suggesting that 
the moduli space is not left invariant 
if we combine FM duality with the shift of B-field and volume. 
 
Further relations between $Z_2$ invariant instantons
on $T^4$ for different ranks and instanton numbers
can be guessed by similar arguments. Here we have 
less powerful statements because we do not have 
very explicit information about the relevant 
moduli spaces. Nevertheless, 
we are led to suggest that the space of $Z_2$ invariant 
instantons for gauge group $U(N)$ with instanton number 
$2k$ on $T^4$ (i.e $k$ on $T^4/Z_2$) has the same 
moduli space as $U(k-N)$ with instanton number $2k$, 
where we have, without loss of generality taken $k > N$.

\newsec{The six-dimensional perspective and the ADS connection} 

Under the headings Rep. I and Rep. II, 
we studied, in sections 4 and 5, two different sectors of the 
theory of zero-branes on $T^4/Z_2$. 
They can be described as two different 
quotients of the theory 
of $N$ four-branes on $\hat T^4$. 
For Rep. II, the group we are quotienting by 
is a $Z_2$ which acts only as reflection $x^a \rightarrow -x^a$ 
on the four spatial coordinates of the $4$-brane. 
For Rep. I we have a $Z_2$ 
quotient, where a spatial reflection 
is combined with a gauge transformation. 
The gauge transformation squares to $1$ and has an equal 
number of $+$ and $-$ eigenvalues. 
In this section it will be convenient to 
choose the gauge transformation as the 
$2N \times 2N$ matrix $ \sigma_1 \otimes 1_{N\times N}$. 
The spatial reflection is, therefore, accompanied 
by exchanging half the branes with the other half. 
 
In this section we will consider the limit
where $\hat T^4$ decompactifies, {\it i.e.}
the limit where $T^4$ goes to zero volume. 
Then we are led to consider the quotient
 of $2N$ four-branes of type IIA on $R^4 \times R^6$, by symmetries 
which involve reflection in $4$ spatial coordinates of $R^4$. 
From the point of view of M-theory 
this is a system of $2N$ five-branes. 

In sections $4$ and $5$ 
we have really described quotients of $4+1$ dimensional 
maximally supersymmetric Yang-Mills theory, which is the 
low-energy description of the theory 
of four-branes. This $4+1$ Yang Mills by itself is 
ill-defined in the UV. One simple way to 
embed it in a well-defined theory 
without gravity is to recognize four-branes 
as five-branes with one leg wrapped along the 
eleventh direction. The strong coupling limit
then decompactifies the extra dimension, 
to give the $(0,2)$ superconformal theory in six dimensions
\refs{\strom,\witcom}. 

The simplest way to make sense of these quotiented
theories at all energy scales is to view them 
as quotients of the $(0,2)$ theory. 
The superconformal fixed point theory should have 
global $U(N)$ symmetry currents with computable correlators,
so there is a well-defined sense in which they have $U(N)$ symmetry.
It therefore makes sense to say that 
we can quotient them in two different ways 
depending on how the $Z_2$ is embedded
in the gauge group. 

We can also define the six-dimensional 
limit as the strong coupling limit of the 
quotiented four-brane theory. 
This theory is not expected to have 
the full six-dimensional Lorentz symmetry
$SO(5,1)$ which is broken by the $Z_2$ 
quotient. However it is expected to have six-dimensional
{ \it scale symmetry} because it is being claimed
to be the six-dimensional fixed point theory that the five-dimensional 
theory flows to. A discussion of the relation between 
scale and conformal invariance appears in \jps\ 
but it is not easy to compare it with this situation 
directly since, first, $SO(5,1)$ is broken, 
and secondly, we do not have an explicit Lagrangian 
description of the general $(0,2)$ theory. 

A related way to understand the six-dimensional 
limit is to recognize that the momentum dual 
to the extra coordinate is just instanton number, 
as exploited in the context of Matrix Theory in \mroz.
We can expect that quantum mechanics on the moduli 
spaces of appropriate $Z_2$ invariant instantons 
will provide the Matrix Model definitions of these
theories along the lines of \refs{\abkss,\witcfth}. 
 Since the large $N$ limit of the five-brane theory 
is conjectured to be related
to M theory on $ADS_7 \times S^4$, it is illuminating 
to look for quotients of this theory which correspond
to the quotients in Rep. I and Rep. II. Quotients 
of ADS backgrounds have been considered 
in \ks\ in order to obtain information about 
 string vacua with reduced supersymmetry. 
There the quotient acts on the sphere. 
Both quotients of interest here act on $ADS_7$ and break
the $SO(6,2)$ conformal symmetry to 
$SO(4) \times SO(2,2)$. 
However there are important differences. 
We recall that ADS is a hyperboloid in an 8-dimensional
embedding space 
\eqn\adssev{  X_0^2 -  \sum_{ i = 1 }^{6 } X_i^2 + X_{7}^2 = 1 } 
The metric on this space can be written in horospheric coordinates,
(see for example \gibb)
appropriate to the five-brane, as
\eqn\hor{ ds^2 = {1\over z^2} (-dt^2 + dz^2 + dx\cdot dx) }  
The quotient related to Rep. II 
acts simply as reversal of the 
4 coordinates parallel to the four-brane. This is a reflection of $x^a$
on the five-brane for $i =1 \cdots 4$, which implies $X^a\to
-X^a$. The fixed surface in this case is thus defined by the
conditions $X^a = 0 $, which gives a hyperboloid, which is
$ADS_3$. The hyperboloid intersects infinity on $S^2$. 

We propose that the quotient related to Rep. I acts as
this reflection together with a reflection in $z$. 
This is motivated by an interesting picture
of `where the brane lives' on ADS space.
Gibbons \gibb\ observes that the metric 
written in horospheric coordinates~\hor\
suggests an interpretation of the ADS 
solution as infinitely many branes aligned 
along the $z$ coordinate. 
As we discussed at the beginning of this section, 
Rep. I corresponds to a quotient where 
a spatial reflection is accompanied 
by exchanging half the branes with the other half.
So the natural proposal is that, on the ADS side, 
the reflection in $x_i$ is accompanied 
by a reflection in $z$. In the hyperboloid coordinates this is
equivalent to $X^0,X^5,X^6,X^7 \to -X^0,-X^5,-X^6,-X^7$. 
There are no fixed surfaces
in the case of this quotient. 

We present some pieces of evidence 
supporting this identification.
First, it leaves unbroken the subgroup
of $SO(4) \times SO(2,2)$. 
In terms of the embedded hyperboloid, the $SO(4)$ rotates $X_1 \cdots
X_4$ while the $SO(2,2)$ rotates the coordinates
$X_0, X_5, X_6, X_7$. 
The $SO(2,2)$ contains scale invariance 
which we expect from the argument
we gave above. 
We can also decompose the $SO(6,2)$ algebra 
in terms of the six-dimensional Lorentz group 
$SO(5,1)$, with generators $M_{\mu \nu} $
(rotations and boosts), 
$P_{\mu}$ (momenta), $K_{\mu}$ (special conformal 
transformations), $D$ (dilatation). Then the surviving 
generators are $M_{ab}$ which generate $SO(4)$,
and $K_0, K_5, P_0, P_5, M_{05}, D$. 
It is easy to check that from the $SO(6,2)$ 
relations that these two sets commute and
that the second set is isomorphic
to $SO(2,2)$, the conformal algebra in $1+1$ 
dimensions. There is still the $Spin(5)$ 
R-symmetry. So the full bosonic 
symmetry group is $SO(4) \times SO(2,2) \times Spin(5)$. 

The $SO(6,2)$ spinors decompose into $SO(5,1)$ spinors as 
$$\pmatrix {Q \cr S^c} $$
where $S^c = C (S^{\dagger} \Gamma_0)^T$. 
$Q$ and $S$ have opposite $SO(5,1)$ chirality. 
The reflection of $X^0,X^5,X^6,X^7$ projects the spinors onto spinors
of definite chirality under $\Gamma^{0}\Gamma^5\Gamma^6\Gamma^7$  
where these are $SO(6,2)$ $\Gamma$ matrices. Since the $SO(6,2)$
spinors have definite $SO(6,2)$ chirality, their chirality under the
$SO(2,2)$ operator $\Gamma^{0}\Gamma^5\Gamma^6\Gamma^7$, is correlated
with their chirality under $\Gamma^1\Gamma^2\Gamma^3\Gamma^4$. 
Thus, since the surviving supersymmetries have definite $SO(2,2)$ 
chirality, they have definite $SO(4)$ chirality. Consequently, 
the number of supercharges in the algebra is $16$. 

We also observe that restricting the
$Q$ and $S$ supercharges to positive 
$SO(4) $ chirality is consistent with the 
the truncation of the bosonic generators to those 
given above. The anti-commutators
between $Q$ or between $S$ in the six-dimensional superconformal
algebra, are given for example in \kal, 
\eqn\consrest{\eqalign{  
   & \{ Q^{i}_{\alpha^{\prime}} , Q_{\beta^{\prime}}^j  \} 
      = -2 (\gamma_{\mu})_{\alpha^{\prime} \beta^{\prime}}
     \Omega^{ij} P^{\mu} \cr 
   &  \{ S^{i}_{\alpha}  , S_{\beta}^j  \} 
      = -2 (\gamma_{\mu})_{\alpha \beta }
      \Omega^{ij} K^{\mu} \cr 
}}
One finds that when the spinors are of the same $SO(4)$ chirality, 
$P_0$, $P_5$, $K_0$, $K_5$ can occur on the right 
hand side, but $P_i$, $K_i$ cannot. 

There are also non-vanishing commutators of the form 
\eqn\com{  
   \{ Q^i_{\alpha'} , S^j_\beta \} =  
      -2 c_{\alpha'\beta}\left(\Omega^{ij}D+4U^{ij}\right)
      - 2 (\gamma_{\mu\nu})_{\alpha'\beta}\Omega^{ij}M^{\mu\nu} } 
The $SO(4)$ chirality conditions on $Q$ and $S$, then imply that only
 $M_{05}$ and $M_{ab}$ appear.
 If we look at the bosonic subalgebra, we have
a product of three simple factors $SO(4) \times SO(2,2) \times
Spin(5)$. If we look at the full superalgebra, however, these factors
are not decoupled. The algebra is not simple. We know it
cannot be, because a superalgebra with such a product as bosonic
part does not appear in Kac's classification of Lie superalgebras
\kac\ (a recent paper containing a review 
of some of the relevant facts is \min.). 
 
The $SU(2)_R$ factor of $SO(4)$ and 
a similar $Sl(2)_R$ factor of $SO(2,2)$ 
act trivially on the spinors since we have
spinors of a definite chirality under these two groups.
So we can rewrite the algebra in terms of
a decoupled $SU(2)_R \times Sl(2)_R $ part, 
together with a superalgebra whose 
bosonic part is $SU(2)_L\times Sl(2)_L \times Spin(5)$. 
So this is an interesting scale invariant fixed 
point in six dimensions
which contains a product made of a bosonic algebra 
$SU(2)_R\times Sl(2)_R$ and a superalgebra, with $16$ 
supercharges and a bosonic part  
$SU(2)_L\times Sl(2)_L \times Spin(5)$
as symmetry group. 
It will be interesting to give a more complete
description of such superalgebras, containing 
scale invariance, and their relevance to  
higher dimensional branes, {\it e.g.}
of the kind discussed in \hull. 

The unbroken bosonic symmetry group is the
same in Rep. I and Rep. II. 
Furthermore, the surviving supersymmetry in both cases are
also the same. In Rep. I we saw that the reflection projects onto
spinors of positive chirality under
$\Gamma^{0}\Gamma^{5}\Gamma^{6}\Gamma^{7}$. Since they had positive
$SO(6,2)$ chirality, this implied they had positive chirality under
the $SO(4)$ operator $\Gamma^{1}\Gamma^{2}\Gamma^{3}\Gamma^{4}$. For
Rep. II, the reflections projects onto spinors of positive chirality
under $\Gamma^{1}\Gamma^{2}\Gamma^{3}\Gamma^{4}$. Again this implies
the supercharges have positive $SO(4)$ and $SO(2,2)$ chirality, and we
are left with the same supersymmetry algebra in each case. 
From the point of  
view of the five-brane world-volume theory, 
we are looking at different ways of embedding the $Z_2$ in the gauge
group, which should not affect the surviving spacetime
symmetries. This is analogous to the 
fact that expanding 
around different Wilson line backgrounds does not 
change the SUSY algebra.  
 
Note that preserving the 
$SO(4) \times SO(2,2)$ symmetry 
constrains the class of scale invariant 
fixed points. The B-field 
at the fixed point which is a 
parameter in the weak coupling
limit of the D-brane theory is no longer a 
physical parameter in the strong coupling limit.
In other words, it is a perturbation of the four-brane 
theory which becomes irrelevant in the strong coupling
limit.  
The blow-up parameters will break the $SO(4)$ symmetry.    
Thus, the only freedom we have is to add membrane charge. The ADS
picture suggests that the description in terms of supergravity will
become harder, with a more complicated fixed surface. 

In the above discussion of Rep. I, we combined 
the spatial reflection with 
a symmetry $J$ which represented the 
a non-trivial gauge transformation. It is  in the centre of $SO(6,2)$, 
and satisfies $J^2 = 1$, and can be understood
geometrically by viewing the hyperboloid as a double cover. 
To obtain generators of $Z_k$ satisfying $J^k = 1$,
as may be appropriate 
for more general ALE spaces, 
we may, as in \gibb\ 
consider multiple covers.

\newsec{Summary and conclusions} 

We have studied a class of representations
of a non-commutative algebra in relation to 
branes on $T^4/Z_2$. The regular rep (Rep. I) 
described zero-branes on the orbifold. 
We checked that it gave the correct moduli 
spaces of vacua for both the Higgs and Coulomb branches. 
We have also found a description of the regular rep
in terms of a quotient of the theory of four-branes
on the dual torus, $\hat T^4$, which 
corresponds to a non-trivial bundle on $ \hat T^4/Z_2$ 
(with singularities at the fixed points). 
 
We also studied a class of
representations corresponding 
to the addition of two-branes at different 
fixed points of $T^4/Z_2$. 
As for the regular rep, we checked that the Coulomb 
and Higgs branches are as expected, and 
gave a formulation in terms of bundles on $\hat T^4/Z_2$. 

The discussion of bundles on $\hat T^4/Z_2$ 
involves two algebras. The first one is obvious, it is just 
the {\it commutative} algebra 
of functions on $\hat T^4/Z_2$, i.e even 
functions on the covering $\hat T^4$. 
There is another {\it non-commutative } algebra, 
the group algebra of the quotienting group
which defines $T^4/Z_2$, which
played an important role in describing 
the physics of branes and bundles on $\hat T^4/Z_2$.
 
Whereas, on tori, either commutative or 
non-commutative algebras can lead, via their regular
rep, to 
a Higgs branch moduli spaces of $ S^N(T^4)$, 
(depending on whether we have turned on B-fields or not), 
our examples strongly suggest that 
only a non-commutative algebra leads to 
moduli spaces of the type $S^N (T^4/Z_2)$. 
A second important role for the 
non-commutative algebra is that 
it provides operators whose traces 
are related to brane charges at the fixed
point, as touched upon at the end of section six. 

We found the relation between 
the duality of the brane description
on $T^4/Z_2$ to the one on $\hat T^4/Z_2$
and conventional T-dualities. We wrote 
down the relevant element of $O(\Gamma_{4,20})$. 
Interestingly it mixes twisted sector
states with untwisted sector ones. This should 
be understandable using boundary states\foot{We are grateful
for an illuminating discussion with Ashoke Sen on this subject.} 
along the lines of \sen. 
We deduced some properties of 
$Z_2$ symmetric instantons on $T^4$ using 
K3 dualities.

We observed that quotients
of four-brane theories are very useful
in describing the physics of branes on the 
orbifold. Regarding these as $5$-branes 
in M theory, we identified some of the 
relevant quotients on the ADS side. 
The identification of the 
ADS background related to Rep. II is straightforward, 
The identification of the background corresponding
to Rep. I is less straightforward, because
the spatial reflection is accompanied by 
an action on the gauge indices labeling the 
brane. We proposed an identification based
on the picture \gibb\ that the branes are stacked along 
the $z$-coordinate \hor. According to this picture, the 
answer to the question of where in ADS space 
the branes live is, colloquially, 
`everywhere'. We analyzed the surviving bosonic and fermionic 
symmetries of the quotient and argued that they are expected 
physically. 

\bigskip

\noindent{\bf Note added:}
After this paper was submitted, 
a related paper \grlapi\ appeared, having some  overlap 
with our discussion of the moduli space of zero-branes
and including an interesting discussion of blow-ups of 
the orbifold. 

\bigskip

\noindent{\bf Acknowledgments} 

\nobreak
It is a pleasure to acknowledge pleasant and useful discussions
with O. Aharony, P. Berglund, O. Ganor, R. Gopakumar, 
Z. Guralnik, P.M. Ho, S. Kachru, D. Lowe, 
S. Paul and A. Sen. This work was supported by DOE Grant DE-FGO2-91-ER40671. 
SR thanks the ITP for hospitality while part of this work was done, 
and participants of the duality program for useful discussions. 

\appendix{A}{Review of and remarks on the dualities of K3 
compactifications of type IIA strings} 

In this appendix we summarize the moduli space of type IIA strings on
K3 following Aspinwall~\Kt. In addition, we derive a mass formula for
the BPS brane configurations, and discuss duality in the presence of
$B$-field. Most of the discussion is for a general K3, but in the
final part we specialize, deriving conditions on the form of
a duality transformation which preserves the perturbative $T^4/Z_2$
orbifold, up to inverting the volume. 

The total integer cohomology
$H^*(K3,Z)=H^0(K3,Z)+H^2(K3,Z)+H^4(K3,Z)$
forms a lattice $\Gamma_{4,20}$. Elements 
of the lattice are associated to cycles
as explained in~\Kt. The inner product $u\cdot v$ 
between two elements of the lattice comes
from the oriented intersection number of the corresponding cycles,
\eqn\innerprod{
   u \cdot v = \int_{K3} u \wedge v
}
Thus, for instance, if $\omega$ generates $H^0(K3,Z)$ and
$\omega^*$ generates $H^4(K3,Z)$ we have 
\eqn\omegarels{
   \omega \cdot \omega = \omega^* \cdot \omega^* = 0 \qquad
   \omega \cdot \omega^* = 1
}
 It is natural to decompose the lattice into
the orthogonal sublattices $H^2(K3,Z)$ and $H^0(K3,Z)+H^4(K3,Z)$, 
\eqn\latticedecomp{
   \Gamma_{4,20} = \Gamma_{3,19} + \Gamma_{1,1}.
}

The gravitational and
$B$-field moduli of the string theory are fixed by giving a spatial
four-plane $\Pi$ in the total cohomology over the reals,
$H^*(K3,R)\cong R^{4,20}$. Note that $\Gamma_{4,20}=H^*(K3,Z)\subset
H^*(K3,R)$ is a lattice in $R^{4,20}$. The Einstein metric and
$B$-field can be extracted from the plane in the following way.

First one decomposes $R^{4,20}=R^{3,19}+R^{1,1}$ into $H^2(K3,R)$ and
$H^0(K3,R)+H^4(K3,R)$ as was done for the lattice in
\latticedecomp. Since $R^{3,19}$ has only three spacelike directions,
the projection $\Sigma$ of $\Pi$ into $R^{3,19}$ must be a spatial
three-plane. This plane fixes the Einstein metric on K3 up to the
overall volume. It can be described in terms of three orthonormal
elements of $H^2(K3,R)$, which we will call $s_i$ with
$i=1,2,3$. These components represent the real and imaginary parts of
the complex structure and the K\"ahler form scaled so that volume of
the K3 is one. Making rotations between the $s_i$ defines the same
plane and this represents the freedom to rotate between the sphere of
complex structures $S^2$ on the same hyper-K\"ahler manifold. To
define the full four-plane we need to give an additional vector lying
at least partly in the $R^{1,1}$ space. Aspinwall \Kt\ introduces the
vector 
\eqn\Bprime{
   B' = \omega^* + \alpha\omega + B
}
normalized such that $\omega\cdot B'=1$ and where $B\in H^2(K3,R)\cong
R^{3,19}$. For $B^{\prime} $ to be spacelike, 
\eqn\Bpcond{
   B'\cdot B' \equiv 2V = 2\alpha + B\cdot B > 0, 
}
where the positive quantity $B.B + 2 \alpha$ has been 
identified with the volume $V$ of the $K3$ manifold,
 while $B\in
H^2(K3,R)$ then gives the value of
the $B$-field in string units. Evidence for this identification will
come from calculating the masses of various BPS brane states.
The full four-plane $\Pi$ is then described by the span of the set of
orthogonal vectors
\eqn\Pidef{
   B', \qquad {\hat s}_i = s_i - (B'\cdot s_i)\omega
}

The actual moduli space is a little smaller than simply choosing a
four-plane in $R^{4,20}$. One must identify planes which are related
by the discrete group of rotations which preserve the lattice
$O(\Gamma_{4,20})$.

The lattice also provides a simple labeling of BPS
states. From heterotic--type IIA duality, on the heterotic side, a 
given point in the lattice is a supersymmetric state of fixed 
winding and momentum. Under duality,
these should map to supersymmetric states in the IIA theory. Since on
the IIA side the lattice points are elements of the integer
cohomology, the corresponding BPS states are naturally combinations of
four-branes, membranes and zero-branes wrapping the different cycles
of the K3. In general we can write a point in the lattice as 
\eqn\pvec{
   p = Q_0 \omega + Q_2 + Q_4 \omega^*
}
where $Q_2 \in H^2(K3,Z)$. We will find 
that $Q_4$ is equal to the number of four-branes in the BPS state and
$Q_2$ gives the number of membranes wrapping the two-cycles of the
K3. However $Q_0$ equals the number of zero-branes minus the
number of four-branes. This shift is due to the zero-brane charge
induced by the presence of four-branes due to the curvature of the K3. 
In other words, $p$ is the Mukai charge. 

 From the heterotic point of view, the projection of $p$ onto the plane
$\Pi$ gives the momentum of the state in the (conventionally
right-moving) supersymmetric sector, while the component of $p$
perpendicular to the plane gives the momentum in the bosonic
sector. From this we see that the square of the projection of $p$ onto
$\Pi$ should give the mass of the BPS state in the heterotic string
frame. Thus, as a test of the interpretation
of the K3 moduli space, we calculate this mass on
the type IIA side. This can then be compared with the known expression
for the total Ramond-Ramond (RR) charges of a collection of
branes. Projecting onto $\Pi$ and squaring we find 
\eqn\branemass{
   m^2 =  { 1\over V}  \left\{ 
            \left[ Q_4V + Q_0+Q_2\cdot B -\hf Q_4 B\cdot B \right]^2
            + V\sum_i\left[Q_2\cdot s_i - Q_4 B\cdot
               s_i\right]^2 \right\}
}

The above formula has been written with a special 
choice of units. To make everything explicit,
we should understand 
\eqn\Vdef{
   V = {V_{K3} \over \left(4\pi^2\alpha'\right)^2}
}
where $V_{K3}$ is the physical volume of the K3.
Further, there is an overall factor of 
$ l_{het}^{-2} = {e^{-2\phi_6} \over \alpha'V} $. 
We also have the relation between the ten-dimensional dilaton $\phi$
and $\phi_6$ as $e^{-2\phi_6}=Ve^{-2 \phi }$. 

We can then read off the amount of four-, two- and zero-brane charge
from their respective contributions to the mass. We find
\eqn\cench{\eqalign{
   q_4 &= Q_4 \cr
   q_2 &= Q_2- Q_4 B \cr
   q_0 &= Q_0 + Q_2\cdot B -\hf Q_4 B\cdot B
}}
This can be compared with the gauge theory description of the physical
RR charges in the presence of $B$-field and on a curved manifold. With
the normalization of $B$ given above, we have~\ghm, for a $U(N)$ 
bundle,
\eqn\WZ{
   q = \tr\,e^{i\left(F/2\pi-B\right)}\sqrt{{\hat A}(K3)} 
}
so that 
\eqn\RRcharge{\eqalign{
   q_4 &= N \cr
   q_2 &= {i\over 2\pi}\tr F - iN B \cr
   q_0 &= - {1\over 8\pi^2}\tr F\wedge F - N 
              + {1\over 2\pi}\tr F\wedge B - \hf N B\wedge B
}}
where we have used $p_1=48$ for a K3. We see the dependence
on the $B$ field exactly matches~\cench, and $Q_4=N$, $Q_2=i\tr
F/2\pi$, while $Q_0=-\tr F\wedge F/8\pi^2-N$. The shift in the last
expression corresponds to the induced zero-brane charge due to the
curvature of the K3. The inclusion of this term justifies 
the identification of 
the $Q$'s with the Mukai charges. Note that the contribution to the RR
charge due to the $B$-field, which is generally non-integer, is not
included in the description of the state as a point in the lattice. The
$B$-field contributions are due to the projection of $p$ onto $\Pi$,
and for the $\hf B\cdot B$ term, because we defined the volume as
$\alpha+\hf B\cdot B$.
  
We now turn to the question of duality in the presence of
$B$-field. We have seen in section 3 that Fourier-Mukai duality
without $B$-field corresponds to the element of $O(\Gamma_{4,20})$
which exchanges of $\omega$ and $\omega^*$, sending $V$ to $1/V$.
For simplicity, let us assume that $ B\cdot s_i = 0 $. 
Then the Fourier-Mukai transformation 
\eqn\fmtransf{
   \omega \to \bom = \omega^* \qquad \omega^* \to \bom^* = \omega
}
gives 
\eqn\bpfm{
   B' = \bom + \alpha \bom^* + B \to 
   {\bar B}' = \bom^* + \bom/\alpha + B/\alpha
}
where we have rescaled so that ${\bar B}'$ has the correct
normalization in the new basis. This means that 
\eqn\fmvol{
   V \to {V \over \left(V - \hf B\cdot B\right)^2} \qquad 
   B \to { B \over {V - \hf B\cdot B}}
}
so that in general the volume is not inverted, and $B$ is rescaled. 

Finally let us specialize to discuss duality transformations for the
$T^4/Z_2$ orbifold limit of a K3. The orbifold has sixteen collapsed
two-cycles $\gamma_i$, one at each fixed point, as well as six
two-cycles inherited from the $T^4$. While these cycles are a
basis for $H^2(K3,Q)$, they are not correctly normalized to generate
$\Gamma_{3,19}$ \asp. Instead they generate a sublattice called the Kummer
lattice. As discussed in section 3, to get a perturbative string
background we must have $B$-flux on each of the sixteen collapsed
two-cycles. In our normalization this gives: 
\eqn\ktbp{
   B' = \omega^* + \alpha \omega 
          - {1\over 4 } \left(\gamma_1 + \cdots + \gamma_{16} \right)
}
Since $\gamma_i\cdot\gamma_j=-2\delta_{ij}$, we have $B\cdot B =
-2$. Thus we immediately see that the Fourier-Mukai duality
transformation given in~\bpfm, does not invert the volume. Instead,
from~\fmvol, we find $V\to V\left(V+1\right)^{-2}$ and the flux on
each collapsed cycle becomes $\hf\left(V+1\right)^{-1}$. Since the
coefficient of $\gamma_i$ is no longer $1/4$,
this does not represent a perturbative
string background.

It is natural to ask what kind of transformation could invert the
volume and preserves the $B$-flux on each collapsed
cycle. Consider a transformation which rotates among the $\omega^*$,
$\omega$ and $\gamma_i$ vectors. We assume
\eqn\genbp{
   B' \to {\bar B}' = \bom^* + {\bar\alpha} \bom 
          - { 1 \over 4 } 
        \left({\bar\gamma}_1+\cdots+{\bar\gamma}_{16}\right)
}
such that the volume is inverted 
\eqn\Vinv{
   V = \alpha-1 \to {\bar V} = {\bar\alpha}-1 = A/V
}
with some $A$. That this is true for all $\alpha$ immediately gives
the result 
\eqn\barom{
   \omega = C\left\{ 2 \bom^* +  2 \bom 
       - \hf \left({\bar\gamma}_1+\cdots+{\bar\gamma}_{16}\right)
       \right\}
}
with $C$ undetermined. It might appear that $C$ is required to be a
multiple of two if the transformation is to be symmetry of the
lattice. However, in fact, $\hf
\left({\bar\gamma}_1+\cdots+{\bar\gamma}_{16}\right)$ is an element
of $H^2(K3,Z)$, a result of the Kummer lattice being a sublattice of
$\Gamma_{3,19}$. Thus $C$ is only required to be integer.

\listrefs

\end